\documentclass[a4paper, amsfonts, amssymb, amsmath, reprint, showkeys, nofootinbib, twoside, onecolumn,notitlepage]{revtex4-2}

\def\prd{{Phys.~Rev.~D}}        
\def\prl{{Phys.~Rev.~Lett.}}    
\usepackage{amsmath,amstext}
\usepackage[T1]{fontenc}
\usepackage{amssymb}
\input{epsf}
\usepackage{graphicx}
\usepackage{ae,aecompl}
\usepackage{combelow}

\usepackage{standalone}
\usepackage{hyperref}
\usepackage{amsmath}
\usepackage{amssymb}
\usepackage{mathtools}
\usepackage{bm}
\usepackage{cleveref}
\usepackage{tensor}
\usepackage{braket}
\usepackage{enumitem}
\usepackage{mhchem}
\usepackage{cancel}
\usepackage{tikz}
\usepackage{pgfplots}
\usetikzlibrary{external}
\usepackage{mathrsfs}
\usepackage{amsthm}
\usepackage{upgreek}

\usepackage{color}

\newcommand{\spc}{\quad \quad \quad}

\def\be{\begin{equation}}
\def\ee{\end{equation}}
\def\beq{\begin{eqnarray}}
\def\eeq{\end{eqnarray}}

\theoremstyle{definition}

\theoremstyle{theorem}
\newtheorem{theorem}{Theorem}
\newcommand{\Kn}{\textup{Kn}}
\theoremstyle{corollary}

\begin{document}
\title{Convergence of the hydrodynamic gradient expansion in relativistic kinetic theory}
\author{L.~Gavassino}
\affiliation{Department of Mathematics, Vanderbilt University, Nashville, TN 37211, USA}

\begin{abstract}
We rigorously prove that, in any relativistic kinetic theory whose non-hydrodynamic sector has a finite gap, the Taylor series of all hydrodynamic dispersion relations has a finite radius of convergence. Furthermore, we prove that, for shear waves, such radius of convergence cannot be smaller than $1/2$ times the gap size. Finally, we prove that the non-hydrodynamic sector is gapped whenever the total scattering cross-section (expressed as a function of the energy) is bounded below by a positive non-zero constant. These results, combined with well-established covariant stability criteria, allow us to derive a rigorous upper bound on the shear viscosity of relativistic dilute gases.
\end{abstract}

\maketitle

\vspace{-1cm}

\section{Introduction}

Euler discovered the equations of ideal fluid dynamics in 1752 \cite{Christodoulou2007Compressible}. At that time, the stress tensor of fluids had no gradient terms. Navier and Stokes added the first-order gradient corrections one century later (Stokes completed the formulation in 1850 \cite{HistoryNavierStokes}). Then, it took almost eighty years until Burnett added second-order gradient corrections, in 1935 \cite{Burnett1935}. Nowadays, third-order gradient terms are becoming increasingly popular \cite{El:2009vj,Jaiswal:2013vta,GrozdanovThirdOrder2015kqa,deBrito:2023tgb}. This pattern may continue in the future. The natural question is: Will the approximation error of hydrodynamics approach zero in the limit as the number of derivative corrections tends to infinity?

Let us formulate this question more precisely. Fix a fluid that undergoes some flow of interest (e.g. Couette flow). By changing the typical lengthscale $L$ of the flow, it should be possible to express, e.g., the stress component $\uppi^{12}$ as a function of the Knudsen number $\Kn=\lambda/L$, where $\lambda$ is some characteristic microscopic scale of the fluid (usually the particles' mean free path). It is straightforward to see that the $n^{\text{th}}$-order gradient correction to $\uppi^{12}$ must scale like a power $\Kn^n$ \cite{huang_book}. Thus, the question about the accuracy of hydrodynamics boils down to asking whether there is a neighborhood of $\Kn=0$ where the function $\uppi^{12}(\Kn)$ coincides with its Taylor series, namely \cite{Struchtrup2011}
\begin{equation}\label{serione}
\uppi^{12}(\Kn) \equiv \sum_{n=0}^{+\infty} \uppi^{12}_{(n)} \, \Kn^n \, .
\end{equation}
In the kinetic theory of dilute gases, the series above is known as the ``Chapman-Enskog expansion'' \cite{Groot1980RelativisticKT,cercignani_book,Santos1986}. 

A necessary condition for \eqref{serione} to hold is that the power series on its right-hand side converge. Unfortunately, this rather simple requirement turns out to depend on the details of the flow. It has been argued \cite{Santos1986,Romatschke2017} that, in almost any flow whose primary fluid variables are not entire functions of spacetime (except for a set of measure zero \cite{DenicolNoronhaExactHardSphereses}), the series \eqref{serione} diverges factorially. Considerable evidence supports this claim \cite{HellerAttractors2015,HellerBeyondBjorken2022,HellerSingulant2022,GavassinoInfiniteOrder2024pgl}. On the other hand, it was also shown that, if one works with ``hydrodynamic waves'', namely \textit{sinusoidal} linear fluctuations of conserved densities, then the gradient series converges in strongly coupled systems \cite{GrozdanovGradient2019kge}, and in dilute gases with hard interactions \cite{McLennan1965,Dudynski1989}. 

To prove the above result, the authors of \cite{GrozdanovGradient2019kge,McLennan1965,Dudynski1989} showed that, for sinusoidal linear waves, the calculation reduces to an eigenfrequency problem. To understand how this works, consider the example of a linear shear wave propagating in direction 1 and fluctuating in direction 2. The relevant conserved density for this problem is the component $T^{02}$ of the stress-energy tensor, whose evolution equation reads $\partial_t T^{02}=-\partial_1 \uppi^{12}$. Using equation \eqref{serione}, we can express $\uppi^{12}$ as a linear function of the space derivatives\footnote{Strictly speaking, $\uppi^{12}$ may depend also time derivatives of the fluid variables \cite{GavassinoInfiniteOrder2024pgl}. However, it is possible (by a so-called ``resummation'') to replace each time derivative with an infinite series of space derivatives. The first to systematically apply this technique to the gradient expansion were Chapman and Enskog themselves. The simplest example of resummation is the following. Take the Benjamin-Bona-Mahony equation $\partial_t \varphi=-\partial_x \varphi +\partial_t \partial^2_x \varphi$ \cite{Benjamin1972,GavassinoDispersions2024}, and ``isolate'' the time derivative $\partial_t \varphi=-(1{-}\partial^2_x)^{-1}\partial_x \varphi=-(1{+}\partial^2_x{+}\partial^4_x{+}...)\partial_x \varphi$. It is easy to show, using Fourier analysis, that the result is exact within a finite neighbourhood of $\partial_x \equiv ik=0$.} of $T^{02}$, and look for solution of the form $T^{02}=e^{i(kx_1-\omega t)}$. Then, as claimed, the gradient series \eqref{serione} is converted into a Taylor series for the  eigenfrequency $\omega$ expressed in terms of $k$:
\begin{equation}\label{dispersionwk}
\omega(k)=\sum_{n=1}^{+\infty} b_{n} k^n \, .
\end{equation}

Nearly 40 years ago, \citet{Dudynski1989} found sufficient conditions for the convergence of \eqref{dispersionwk} in relativistic gases. The mathematical analysis of \cite{Dudynski1989} is very technical, and the main result depends on the detailed structure of the Boltzmann integral, making it hard to generalize. Here, we provide a generalization of the analysis of \cite{Dudynski1989}, in the form of a novel and universal statement, valid for both Boltzmann's equation and all (reasonable) approximations thereof. Namely:
\begin{theorem}\label{theo1}
Consider a relativistic kinetic theory for classical particles whose non-hydrodynamic sector has a finite gap, $\omega_g{=}-i/\tau_g$, at $k=0$. Suppose that the linearised collision operator is consistent with thermodynamic principles (see Sec. \ref{assumpzione} for all the details). Then, the Taylor series \eqref{dispersionwk} of any hydrodynamic dispersion relation $\omega(k)$ has a non-vanishing convergence radius. In particular, the radius of convergence of shear waves has the following lower bound:
\begin{equation}\label{TheBound}
 R_{\text{shear}} \geq (2\tau_g)^{-1} \, .   
\end{equation}
\end{theorem}
The physical content of this theorem is rather intuitive: Hydrodynamics exists whenever there is a genuine separation of scales, marked by a characteristic ``microscopic'' lengthscale $\tau_g$. The bound \eqref{TheBound} comes from the fact that, at $|k| \sim (2\tau_g)^{-1}$, the complexified dispersion relation of the shear wave may collide with the first non-hydrodynamic dispersion relation, similarly to what happens in the $\mathcal{N}{=}4$ supersymmetric Yang-Mills example \cite{GrozdanovGradient2019kge}. 

In a recent article \cite{GavassinoGapless2024rck}, we proved that there is no non-hydrodynamic gap if the differential cross section tends to zero at high energies. Hence, in all theories where this happens, $\tau_g=0$, and Theorem \ref{theo1} does not apply. For this reason, we provide here a second result, that is somehow complementary to Theorem 2 of \cite{GavassinoGapless2024rck}, namely: 
\begin{theorem}\label{theo2}
Consider a relativistic kinetic theory for classical particles (governed by Boltzmann's equation), whose differential cross section is $\upvarsigma(s,\theta)$, where $s$ is the Mandelstam variable, and $\theta$ is the scattering angle in the center of momentum frame. Suppose that the total cross-section is bounded below by a positive constant, namely
\begin{equation}\label{assuziuz}
\inf_{s\in \mathbb{R}^+} \int_{\mathcal{S}^2} \upvarsigma(s,\theta) d^2\Omega=\sigma_0 >0 \, .
\end{equation}
Then, the non-hydrodynamic sector has a finite gap at $k=0$.
\end{theorem}
Also this result is quite intuitive. If the total cross section is non-vanishing at all energies, all particles must collide within a finite timescale $\tau_g$, and this timescale defines the size of the gap.

Throughout the article, we adopt the metric signature $(-,+,+,+)$, and work with natural units $c{=}k_B{=}\hbar{=}1$.

\section{General convergence result}\label{generone}

In this section, we prove the first part of Theorem \ref{theo1}, namely that all hydrodynamic modes have a non-zero radius of convergence, provided that the non-hydrodynamic sector is gapped.

\subsection{Statement of the problem}

Our setting is the same as that of \cite{GavassinoGapless2024rck}. In particular, let $f_p$ (with $p=$ ``four-momentum'') be the kinetic distribution function of a classical relativistic gas, which obeys a Boltzmann-type equation $p^\mu \partial_\mu f=\mathcal{C}$, where $\mathcal{C}$ is some collision term that is kept general. Fix some uniform and static equilibrium state $f^{\text{eq}}_p=e^{(\mu-p^0)/T}$, where $T$ and $\mu$ are constants, representing the temperature and chemical potential of this reference state. Consider the decomposition $f_p=f^{\text{eq}}_p(1+\phi_p)$, and linearize in $\phi_p$. Then, the equation of motion reduces to $p^\mu \partial_\mu \phi_p=L\phi_p$, where $L$ is a linear operator. Since we are interested in dispersion relations like \eqref{dispersionwk}, let us enforce the following spacetime dependence: $\phi_p \propto e^{i(k x_1-\omega t)}$, with $\omega,k \in \mathbb{C}$. This results in the algebraic equation
\begin{equation}\label{mainproblem}
(-i\omega p^0+ikp^1)\phi_p=L\phi_p \, .
\end{equation}
For our purposes, it will be convenient to express equation \eqref{mainproblem} using a slightly different notation. First, let us divide both sides by $p^0$, and define a new linear operator $I=-(p^0)^{-1}L$. Next, let us note that $v^1=p^1/p^0$ is the particles' velocity in the direction $1$. Finally, let us introduce $\lambda = i\omega$ and $\chi = ik$. Then, \eqref{mainproblem} takes the following form \cite{McLennan1965}:
\begin{equation}\label{eignevaluprrop}
(I{+}\chi v^1)\phi_p = \lambda \phi_p \, .
\end{equation}
This is an eigenvalue problem for the one-parameter family of operators $I(\chi)=I+\chi v^1$, with corresponding eigenvalues $\lambda(\chi)$. If we replace, in equation \eqref{dispersionwk}, the frequency $\omega$ with $\lambda/i$ and the wavenumber $k$ with $\chi/i$, we find that the radius of convergence of $\omega(k):\mathbb{C}\rightarrow \mathbb{C}$ coincides with the radius of convergence of $\lambda(\chi):\mathbb{C}\rightarrow \mathbb{C}$. Hence, in the following, we will focus on the latter complex function, which is easier to study. In fact, equation \eqref{eignevaluprrop} has the standard form of an eigenvalue problem for an operator $I$ that is perturbed by an operator $v^1$ through a small coupling $\chi$, and we can apply several (rigorous) convergence results from the theory of holomorphically-perturbed linear operators \cite{Kato_book}.

\subsection{Mathematical setting and theorem assumptions}\label{assumpzione}

If we wish to discuss the convergence properties of \eqref{eignevaluprrop} rigorously, we need to specify a convenient space of functions. As in \cite{GavassinoGapless2024rck}, we choose the space $\mathcal{H}=L^2(\mathbb{R}^3,f^{\text{eq}}_p)$, namely the space of square-integrable complex functions with respect to the measure $f^{\text{eq}}_p d^3p$. This is a separable Hilbert space, with inner product
\begin{equation}
    (\phi_p,\psi_p)= \int \dfrac{d^3 p}{(2\pi)^3} f_p^{\text{eq}} \phi^*_p \psi_p \, ,
\end{equation}
and associated norm $|| \phi_p||=\sqrt{(\phi_p,\phi_p)}$. Note that all polynomials in $p^\nu$ belong to $\mathcal{H}$. The key assumption of Theorem \ref{theo1} (i.e. consistency of $I$ with thermodynamics) boils down to three simple requirements on the collision integral \cite{GavassinoGapless2024rck}, namely that, for all $\phi_p,\psi_p \in \mathcal{H}$, the following applies: 
\begin{equation}\label{propertiesfgofI}
\begin{split}
& I\phi_p=0 \spc \text{if and only if} \spc \phi_p\in \text{span}\{1,p^\nu \} \, , \\
& (\phi_p,I\psi_p)^*=(\psi_p,I\phi_p) \, , \\
& (\phi_p,I\phi_p) \geq 0 \, . \\
\end{split}
\end{equation}
These facts can indeed be shown to follow from thermodynamic principles \cite{GavassinoGapless2024rck,GavassinoCasmir2022,RochaGavassinoFluctu:2024afv}. For example, the third condition is a statement of the second law of thermodynamics. Note that, in the first line, only the ``if'' implication can be argued from thermodynamics alone. The ``only if'' implication has been included because Theorem \ref{theo1} assumes the non-hydrodynamic spectrum to be gapped. Indeed, if there were states $\phi_p$ that are not in the form $a+b_\nu p^\nu$, and such that $I\phi_p=0$, this would imply that the fluid can exist in non-thermal configurations that last forever. Such non-thermal configurations should be interpreted as non-hydrodynamic excitations with vanishing gap\footnote{It should be pointed out that, if $I \neq 0$ is Boltzmann's collision integral, the ``only if'' assumption is automatically true \cite[\S IX.1.b]{Groot1980RelativisticKT}.}. In Appendix \ref{proproiuzzione}, we verify that Boltzmann's collision operator indeed fulfills all properties listed in equation \eqref{propertiesfgofI}, see also \cite{cercignani_book}.

Also $v^1$ possesses some relevant properties, which immediately follow from the definitions:
\begin{equation}\label{proeprtiesfogv1}
\begin{split}
& || v^1 \phi_p ||^2 \leq || \phi_p||^2 \, , \\
& (\phi_p,v^1\psi_p)^*=(\psi_p,v^1\phi_p) \, , \\
& (\phi_p,v^1\phi_p) \in \big[-||\phi_p||^2,+||\phi_p||^2 \big] \, . \\
\end{split}
\end{equation}
The second and third lines of \eqref{propertiesfgofI} and \eqref{proeprtiesfogv1} imply that both $I$ and $v^1$ are semibonded Hermitian operators. Hence, we can use the Friedrichs extension \cite{Pedersenbook}, and promote them to self-adjoint operators in $\mathcal{H}$. Furthermore, the first line of \eqref{proeprtiesfogv1} tells us that $v^1$ is actually a bounded operator, with norm $|| v^1||{=}1$.\footnote{Note that the boundedness of $v^1$ is a specific feature of relativity, which forbids any superluminal motion. From the operator-theory perspective, this is a powerful bound, which allows us to establish more stringent results than those of \cite{McLennan1965} in the non-relativistic theory. In fact, in \cite{McLennan1965}, no general bound on the radius of convergence of shear waves could be provided.} For our purposes, this is all we need to know.

\subsection{Quantum analogy}\label{anLuz}

Given that $I$ and $v^1$ are self-adjoint, we can rigorously map the problem \eqref{eignevaluprrop} into a problem of quantum perturbation theory for the Hamiltonian $H_\chi=H_0+\chi V$, where we identify the kinetic profile $\phi_p$ with the quantum eigenstate $\ket{\phi}$, the collision operator $I$ with the unperturbed Hamiltonian $H_0$, the velocity $v^1$ with the perturbation potential $V$, and the eigenvalue $\lambda(\chi)$ with the energy level $E_\chi$. Then, equation \eqref{eignevaluprrop} becomes the time-independent Schr\"{o}dinger equation
\begin{equation}
    (H_0+\chi V)\ket{\phi}=E_\chi\ket{\phi} \, ,
\end{equation}
and we can import all mathematical techniques directly from the quantum mechanics literature. 

In particular, we can apply a well-known result of quantum perturbation theory \cite{KatoAmazing1949}, which states that, if $V$ is bounded (as in our case), then any discrete \textit{isolated} eigenvalue of $H_\chi$ can be expanded as a Taylor series in $\chi$, and it has a non-vanishing radius of convergence. More precisely, let $E_0$ be an isolated discrete energy level of $H_0$, with multiplicity (i.e. degeneracy) $m$. Then, for sufficiently small $\chi$, there are $m$ discrete (not necessarily distinct) energy eigenvalues $E_{\chi,s}$ ($s=1,2,...,m$) of $H_\chi$ which are analytic in an open neighbourhood of $\chi$, and such that $E_{0,n} = E_0$. An analogous statement holds for the corresponding eigenvectors. The interested reader can also see \cite[\S 136]{RieszNagybook}.

\subsection{Proof of convergence}\label{backhere}

We can apply the above mathematical correspondence with quantum perturbation theory to demonstrate the convergence of the gradient expansion. The proof is given below.

From the first line of \eqref{propertiesfgofI}, we see that $\lambda=0$ is a five-fold degenerate point eigenvalue of $I$, with eigenfunctions $a+b_\nu p^\nu$, corresponding to the summational invariants. Under the assumption that the non-hydrodynamic spectrum is gapped, we have that $\lambda=0$ is an \textit{isolated} element of the spectrum. Thus, according to the discussion of section \ref{anLuz}, we can construct five eigenvalue-eigenvector couples $\{\lambda_s(\chi),\phi_{p,s}(\chi)\}_{s=1}^5$ that fulfill \eqref{eignevaluprrop}, i.e. $(I+\chi v^1)\phi_{p,s}(\chi)=\lambda_s(\chi) \phi_{p,s}(\chi)$, and which can be expanded as power series
\begin{equation}\label{gringo!!}
\lambda_s(\chi)=\sum_{n=1}^{+\infty} c_{sn} \chi^n \, , \spc \phi_{p,s}(\chi)=\sum_{n=0}^{+\infty} g_{p,sn} \chi^n \, ,
\end{equation}
with non-vanishing radii of convergence. Recalling that $\lambda=i\omega$ and $\chi=ik$, we can express the above Taylor series for the eigenvalues $\lambda_s$ in the form \eqref{dispersionwk} for $|k|\leq R_s$,
where $R_s>0$ are the corresponding radii of convergence. These analytic dispersion relations are to be identified with the ``hydrodynamic modes'' of the gas. They are 5, corresponding to two sound waves, two shear waves, and one charge-diffusion wave. The first part of Theorem 1 is, therefore, proven.

\section{Radius of convergence of shear waves}\label{shearone}

In this section, we complete the proof of Theorem \ref{theo1}, by deriving bound \eqref{TheBound}.

\subsection{Transverse rotations as unitary operators}

Consider the one-parameter group of rotations around the $x^1$ axis, represented by the orthogonal matrices
\begin{equation}
    \mathcal{R}(\theta) =
    \begin{bmatrix}
1 & 0 & 0 \\
0 & \cos(\theta) & -\sin(\theta) \\
0 & \sin(\theta) & \cos(\theta) \\ 
    \end{bmatrix} \, .
\end{equation}
We can represent these transformations in our Hilbert space $\mathcal{H}$ as unitary operators, whose action is to rotate the physical state, namely $\mathcal{R}\phi_p \equiv \phi_{\mathcal{R}p}$, where the dependence on $\theta$ is from now on understood. Proving that such transformations are indeed unitary is straightforward. In fact, they are well defined on all $\mathcal{H}$, and they are isometries:
\begin{equation}
(\mathcal{R}\phi_p,\mathcal{R}\psi_p)= \int \dfrac{d^3 p}{(2\pi)^3} f_p^{\text{eq}} \phi^*_{\mathcal{R}p} \psi_{\mathcal{R}p} = \int \dfrac{d^3 q}{(2\pi)^3} f_{\mathcal{R}^{-1}q}^{\text{eq}} \phi^*_{q} \psi_{q}=(\phi_p,\psi_p) \, .
\end{equation}
In the second step, we changed the integration variable into $q=\mathcal{R}p$, and used the property $\det \mathcal{R}=1$. In the last step, we used the fact that, since $f^{\text{eq}}_p=e^{(\mu-p^0)/T}$ depends only on $p^0$, it is invariant under rotations, and $f_{\mathcal{R}^{-1}q}^{\text{eq}}=f_{q}^{\text{eq}}$. A second key feature of  $\mathcal{R}$ is that it commutes both with $v^1$ and $I$. Commutation with $v^1$ is obvious:
\begin{equation}
    \mathcal{R}v^1\phi_p= \mathcal{R} \bigg[ \dfrac{p^1}{p^0} \phi_p \bigg] = \dfrac{(\mathcal{R}p)^1}{(\mathcal{R}p)^0}\phi_{\mathcal{R}p} = \dfrac{p^1}{p^0}\phi_{\mathcal{R}p}= v^1 \mathcal{R}\phi_p \, .
\end{equation}
Commutation with $I$ is a consequence of Lorentz invariance. To see this, consider again the equation of motion $p^\mu \partial_\mu \phi_p=L\phi_p$, and focus on perturbations that are homogenous in space. Then, the general solution can be expressed in the form $\phi_p(t)=e^{-It}\phi_p(0)$. Now, consider an observer that is at rest with respect to the background state, but is rotated by an angle $\theta$ around the $x^1$ axis. From their perspective, the state is $\Tilde{\phi}_p(t)=\mathcal{R}\phi_p(t)=\mathcal{R}e^{-It}\phi_p(0)$. However, the background is isotropic, and the theory is Lorentz-invariant. Hence, the gas obeys the same evolution equations in the reference frame of this rotated observer, meaning that $\Tilde{\phi}(t)$ must also be a solution of the equations of motion, i.e. $\Tilde{\phi}_p(t)=e^{-It}\Tilde{\phi}_p(0)$, with initial state $\Tilde{\phi}_p(0)=\mathcal{R}\phi_p(0)$. Combining these facts together, we conclude that
\begin{equation}
\mathcal{R}e^{-It}\phi_p = e^{-It}\mathcal{R}\phi_p \, .
\end{equation}
Since this must hold for all $t$ and $\phi_p$, the identity $\mathcal{R}I=I\mathcal{R}$ follows. In Appendix \ref{ououritazia}, we verify this statement explicitly in the case of Boltzmann's linearised collision integral.

Finally, we note that $\mathcal{R}$ admits an orthonormal eigenbasis. Take, e.g., the states $(f^{\text{eq}}_p)^{-1/2} \Psi^{(n,l,m)}(p)\in \mathcal{H}$, where $\Psi^{(n,l,m)}$ are the eigenfunctions of the Hydrogen atom (diagonalizing the component $L_1$ of the angular momentum). These states generate $\mathcal{H}$, are orthonormal, and diagonalize $\mathcal{R}$, since $\mathcal{R}(f^{\text{eq}}_p)^{-1/2} \Psi^{(n,l,m)}=e^{im\theta}(f^{\text{eq}}_p)^{-1/2} \Psi^{(n,l,m)}$.

\subsection{Proof of the bound}

Given that $\mathcal{R}$ commutes with $v^1$ and $I$, and is diagonalizable, we know that the eigenspaces of $\mathcal{R}$ are invariant under the action of $I(\chi)$ for all $\chi$. In fact, if a certain $\phi_p$ is an eigenvector of $\mathcal{R}$, namely $\mathcal{R}\phi_p = e^{i\varphi} \phi_p$ for some $\varphi \in \mathbb{R}$, then $\mathcal{R}I(\chi)\phi_p=I(\chi)\mathcal{R}\phi_p=e^{i\varphi}I(\chi)\phi_p$, meaning that $I(\chi)\phi_p$ is also eigenvector of $\mathcal{R}$, with the same eigenvalue as $\phi_p$. This implies that the analysis of section \ref{anLuz} is unchanged if, instead of working in the full Hilbert space $\mathcal{H}$, we restrict our attention to an eigenspace $\mathcal{H}_{\varphi}$ of $\mathcal{R}$, relative to the eigenvalue $e^{i\varphi}$.

Now, we recall that, according to the discussion in section \ref{backhere}, the kernel of $I$ is five-dimensional, and all its elements have the form $\phi_p=a+b_\nu p^\nu$. We can expand such states on the following basis:
\begin{equation}
    \{1, \, p^0, \, p^1, \, p^2{+}ip^3, \, p^2{-}ip^3 \} \, .
\end{equation}
Expressed in this basis, $\mathcal{R}(\theta)$ is diagonal, with eigenvalues $\{1,1,1,e^{i\theta},e^{-i\theta}\}$. It follows that only three of the eigenmodes in \eqref{gringo!!} live in the space $\mathcal{H}_0$  (if we take $\theta \notin 2\pi \mathbb{Z}$), corresponding to longitudinal waves (``spin 0''). These three modes are the two sound modes and the charge diffusion mode. The other two modes are transversal (``spin 1''). One mode lives in the space $\mathcal{H}_\theta$, and it corresponds to the shear wave with positive circular polarization. The other mode lives in the space $\mathcal{H}_{-\theta}$, and it corresponds to the shear wave with negative circular polarization. This separation is convenient, since it tells us that the eigenvalue $\lambda=0$ is \textit{non-degenerate}, if we work within the spaces $\mathcal{H}_\theta$ and $\mathcal{H}_{-\theta}$. Thus, we can apply another well-known theorem of quantum perturbation theory \cite{KatoAmazing1949}, which states that, if an isolated non-degenerate eigenvalue $E_0$ of $H_0$ is perturbed by a bounded potential $\chi V$, the series defining the perturbed energy level $E_\chi=\sum_n c_n \chi^n$ converges as long as $||\chi V|| <D/2$, where $D$ the isolation distance of the eigenvalue $E_0$. In our case, $||V||=1$, $E_0=0$, and the isolation distance is the gap magnitude $|\omega_g|=1/\tau_g$. Theorem 1 is, therefore, proved.

\section{Sufficient condition for the existence of the gap}\label{thegappone}

In this last section, we provide a proof of Theorem \ref{theo2}. Given that the latter is a statement about the eigenspectrum at $k=0$, we will work in the homogenous limit throughout, so that the following equation of motion is assumed: $\partial_t \phi_p=-I\phi_p$, whose formal solution is $\phi_p(t)=e^{-It}\phi_p(0)$.

\subsection{Bound on the collision timescale}

Our first task is to estimate the relaxation timescale of the system.
For a particle with momentum $p^\nu$, its mean-free time, i.e. the average timescale over which such particle travels before colliding, obeys the following formula \cite{Strain2010}:
\begin{equation}\label{relxa}
 \dfrac{1}{\tau_p} = \int \dfrac{d^3 p'}{(2\pi)^3} \, f^{\text{eq}}_{p'} \, v_\varnothing  \int_{\mathcal{S}^2} d^2\Omega \, \upvarsigma(s,\theta) \, ,   
\end{equation}
where $s \geq 4m^2$ is the Mandelstam variable and $v_\varnothing \geq 0$ and is M\o ller's ``velocity'',\footnote{Strictly speaking, M\o ller's velocity is not a proper velocity, since $v_\varnothing \in [0,2]$, which can exceed the speed of light. However, it reduces to the ordinary relative speed in the non-relativistic limit.} whose definitions are
\begin{equation}\label{eighteen}
\begin{split}
s={}& -(p^\mu {+} p'^\mu)(p_\mu {+} p'_\mu)=2m^2-2p^\mu p'_\mu \, , \\
v_\varnothing ={}& \dfrac{\sqrt{s(s-4m^2)}}{2p^0 p'^0} \, .\\
\end{split}
\end{equation}
We now show that, if \eqref{assuziuz} holds, $\tau_p$ can be bounded above by a maximum relaxation timescale $\tau_M$ that does \textit{not} depend on the four-momentum $p^\nu$. To this end, let us first note that, in the expression for M\o ller's velocity, we can invoke the following inequality: $\sqrt{s(s{-}4m^2)}\geq \sqrt{(s{-}4m^2)^2}=s-4m^2$, and we can write $s$ explicitly using \eqref{eighteen}. Furthermore, we note that, in equation \eqref{relxa}, the only quantity that depends on $\theta$ is $\upvarsigma(s,\theta)$. Thus, we can carry out the integral over the solid angle, and bound it below using assumption \eqref{assuziuz}. The result is (notation: $v^j=p^j/p^0$)
\begin{equation}\label{boundone}
\dfrac{1}{\tau_p} \geq \sigma_0 \int \dfrac{d^3 p'}{(2\pi)^3} \, f^{\text{eq}}_{p'} \, \bigg(1-v^j v'_j -\dfrac{m^2}{p^0 p'^0} \bigg) \geq \sigma_0 \int \dfrac{d^3 p'}{(2\pi)^3} \, f^{\text{eq}}_{p'} \, \bigg(1 -\dfrac{m}{p'^0} \bigg) \equiv \dfrac{1}{\tau_M} >0  \, .  
\end{equation}
In the second step we used the fact that, since the equilibrium distribution is isotropic, the term proportional to $v'_j$ averages to zero. Furthermore, we have minimized the last term in the round brackets by replacing $p^0$ with $m$. The resulting quantity $\tau_M$ does not depend on $p^\nu$, and it constitutes the upper bound on $\tau_p$ we were looking for.

\subsection{Free energy decay}

The bound $\tau_p \leq \tau_M$ can be used to derive a powerful estimate on the decay of the free energy due to dissipation.
First, let us introduce some notation. Let $\delta J^0$, $\delta T^{0\nu}$, and $TE^0$ be the perturbations to respectively particle density, four-momentum density, and free-energy density. In the linear regime, the following identities hold \cite{GavassinoCausality2021,GavassinoGapless2024rck,RochaGavassinoFluctu:2024afv}:
\begin{equation}
\begin{split}
\delta J^0 ={}& (1,\phi_p) = \int \dfrac{d^3 p}{(2\pi)^3} f^{\text{eq}}_p \phi_p \, , \\
\delta T^{0\nu} ={}& (p^\nu,\phi_p) = \int \dfrac{d^3 p}{(2\pi)^3} f^{\text{eq}}_p p^\nu \phi_p \, , \\
2E^0 ={}&  (\phi_p,\phi_p) =  \int \dfrac{d^3 p}{(2\pi)^3} f^{\text{eq}}_p |\phi_p|^2 \geq 0 \, . \\
\end{split}
\end{equation}
Recalling that, for homogeneous states, the equation of motion is $\partial_t\phi_p=-I\phi_p$, the facts below follow from \eqref{propertiesfgofI}:
\begin{equation}\label{eetgrthy}
\begin{split}
\partial_t\delta J^0 ={}& -(1,I\phi_p)=-(\phi_p,I1)^*=0 \, , \\
\partial_t \delta T^{0\nu} ={}& -(p^\nu,I\phi_p) = -(\phi_p,Ip^\nu)^*=0\, , \\
\partial_t E^0 ={}&  -(\phi_p,I\phi_p) \leq 0 \, . \\
\end{split}
\end{equation}
The first two lines are the usual conservation laws, expressed in the homogeneous limit. The third line is the statement that the free-energy density $TE^0$ is a non-increasing function of time. Moreover, it can be shown \cite{Strain2010,Speck2011} that, if $I$ is Boltzmann's collision operator, there exists a constant $G>0$ (independent of $\phi_p$) such that\footnote{Note that, in references \cite{Strain2010,Speck2011}, the authors define the linear perturbation differently from us. In fact, their linear degree of freedom is not $\phi_p$, but $\Phi_p=(f^{\text{eq}}_p)^{1/2} \phi_p$. However, our mathematical framework is equivalent to theirs. In fact, their inner product $\langle *,*\rangle$ is related to our inner product $(*,*)$ by the identity $\langle \Psi_p,\Phi_p\rangle =(\psi_p,\phi_p)$. Furthermore, their linear collision operator $\mathcal{L}$ is related to our operator $I$ by the identity $\mathcal{L}=(f^{\text{eq}}_p)^{1/2} I (f^{\text{eq}}_p)^{-1/2}$, from which we obtain $\langle \Phi_p,\mathcal{L}\Phi_p\rangle=(\phi_p,I\phi_p)$.}
\begin{equation}\label{coercitiy}
    (\phi_p,I\phi_p) \geq G (\phi_p,\tau_p^{-1}\phi_p) \spc \text{if } \, \, \, \delta J^0=\delta T^{0\nu}=0 \, .
\end{equation}
Combining this inequality with bound \eqref{boundone}, the third line of \eqref{eetgrthy} becomes
\begin{equation}
\partial_t E^0 \leq -\dfrac{2G}{\tau_M} E^0 \spc \text{if } \, \, \, \delta J^0=\delta T^{0\nu}=0 \, .
\end{equation}
With the aid of the integrating factor $e^{2Gt/\tau_M}$, this inequality becomes $\partial_t(E^0e^{2Gt/\tau_M})\leq 0$, which implies
\begin{equation}\label{chebellalainfo}
    E^0(t) \leq e^{-2Gt/\tau_M}E^0(0) \spc \text{if } \, \, \, \delta J^0=\delta T^{0\nu}=0 \, .
\end{equation}
Hence, we have that, if a homogeneous perturbation conserves the four-momentum and the particle number, its free energy must decay no slower than exponentially. Note that equation \eqref{chebellalainfo} implies that all the moments $\delta \rho^{0\nu_1 ... \nu_l}$ of the perturbation decay at least exponentially. In fact, recalling that $||\phi_p||^2=2E^0$, we can use the Cauchy-Schwartz inequality to derive the following bound:
\begin{equation}
|\delta \rho^{0\nu_1 ... \nu_l}(t)|=|\big(p^{\nu_1}...p^{\nu_l},\phi_p(t)\big)| \leq || p^{\nu_1}...p^{\nu_l} || \, || \phi_p(0) ||e^{-Gt/\tau_M} \spc \text{if } \, \, \, \delta J^0=\delta T^{0\nu}=0 \, .
\end{equation}

\subsection{Proving the existence of the gap}

Adopting the same notation of \cite{GavassinoGapless2024rck}, let us introduce the Hilbert subspace $\mathcal{H}_{\text{non-hy}}=\{1,p^\nu \}^{\perp}$. This is the space of states $\phi_p$ such that $\delta J^0=(1,\phi_p)=0$ and $\delta T^{0\nu}=(p^\nu,\phi_p)=0$. Recalling that we are working in the homogeneous limit, these states describe pure deviations from local thermodynamic equilibrium, and should be interpreted as the non-hydrodynamic excitations of the gas. Note that $\mathcal{H}_{\text{non-hy}}$ is an invariant space of $I$, since $I\phi_p$ always belongs to $\mathcal{H}_{\text{non-hy}}$, by the first two lines of equation \eqref{eetgrthy}. Thus, we can restrict $I$ to a self-adjoint operator on $\mathcal{H}_{\text{non-hy}}$. Therefore, combining again equations \eqref{coercitiy} and \eqref{boundone}, and invoking Theorem 2.19 of \cite[Chapter 2, \S 2.4]{Teschlbook}, we finally obtain a lower bound on the non-hydrodynamic gap magnitude $\tau_g^{-1}$, namely
\begin{equation}
\dfrac{1}{\tau_g} = \inf_{\mathcal{H}_{\text{non-hy}}} \text{Spectrum}[\, I \, ] = \inf_{\mathcal{H}_{\text{non-hy}}} \dfrac{(\phi_p,I\phi_p)}{(\phi_p,\phi_p)} \geq \dfrac{G}{\tau_M} >0 \, .
\end{equation}
Thus, the non-hydrodynamic sector has a non-vanishing gap, which is of the order of the timescale over which the free energy decays to zero, see equation \eqref{chebellalainfo}. This completes our proof of Theorem 2.

\section{Final Remarks}

Through Theorem \ref{theo1}, we have provided sufficient conditions for the convergence of the dispersion relations $\omega(k)$ of all hydrodynamic modes in relativistic kinetic theory. The theorem is applicable also in the relaxation-time approximation (see \cite[\S 4.1]{Brants:2024wrx} for an explicit verification), since the latter fulfills all the Theorem's assumptions. It can be shown that, if the assumptions of Theorem \ref{theo1} are met, also the following facts naturally follow, as direct consequences of our proof:
\begin{itemize}
\item We can expand in power series not only the eigenfrequencies $\omega(k)$, but also the corresponding eigenstates $\phi_p(k)$ of the distribution function \cite{KatoAmazing1949,Dudynski1989,McLennan1965}, and their radii of convergence are also finite.
\item The functions $\omega(k)$ and $\phi_p(k)$ coincide with their Taylor series where the latter converge \cite{KatoAmazing1949}.
\item The full Chapman-Enskog expansion converges to the correct kinetic theory results in the linear regime in a neighborhood of $\Kn{=}0$ \cite{McLennan1965}.
\end{itemize}
In conclusion, if the non-hydrodynamic sector has a finite gap, the gradient expansion has solid mathematical foundations within relativistic kinetic theory. Adding up all the infinite terms of the gradient series takes us, from linearized hydrodynamics, back to the linearized Boltzmann equation (in a neighborhood of $\Kn=0$). We stress that, while Theorem \ref{theo1} assumes that the number of particles is conserved, the main result straightforwardly generalizes to situations where creation processes are allowed (e.g., for neutral pions). In this case, the hydrodynamic modes become 4 (instead of 5), the function $1$ no longer appears in the first line of \eqref{propertiesfgofI}, and one needs to set $\mu=0$ in equilibrium. 

Theorem \ref{theo1} also provides a lower bound to the radius of convergence of shear waves in terms of the relaxation time $\tau_g$ of the slowest non-hydrodynamic mode\footnote{A similar strategy as the one used in section \ref{shearone} may be applied to derive analogous bounds for the sound modes and the charge diffusion mode. However, we could not manage to find a unitary operator that plays the same role as $\mathcal{R}$.}. This result formalizes the widespread intuition that the hydrodynamic series can only begin to diverge close to the non-hydrodynamic scale \cite{HellerHydrohedron2023jtd}. Since a similar result is known to hold within $\mathcal{N}=4$ supersymmetric Yang-Mills theory and Israel-Stewart theory, it is natural to wonder whether a more general proof exists. Indeed, we already know that, in dissipative systems, the Taylor series of $\omega(k)$ must break down at some point, by Corollary 1.1 of \cite{HellerBounds2022ejw}. The only question is whether the scale at which this happens must always be related to $\tau_g$. This issue will be addressed in an upcoming paper.  

The knowledge of the radius of convergence of shear waves allows one to derive an infinite number of bounds on transport coefficients \cite{HellerHydrohedron2023jtd}. For example, if we combine our equation \eqref{TheBound} with Theorem 4 of \cite{HellerBounds2022ejw}, we obtain the following rigorous (and universal) upper bound on the shear viscosity $\eta$ computed from kinetic theory\footnote{This bound is similar to the well-known inequality $\eta \leq 5(\varepsilon{+}P)\tau_\pi$, which also comes from kinetic theory \cite{Ghiglieri:2018dgf}. However, we stress that $\tau_g$ and $\tau_\pi$ are different quantities (if the latter is defined from the Kubo formula \cite{Denicol_Relaxation_2011,WagnerGavassino2023jgq}), and the two bounds are not equivalent.}:
\begin{equation}
\eta \leq \dfrac{32}{3\pi} (\varepsilon{+}P) \tau_g \, ,
\end{equation}
where $\varepsilon{+}P$ is the relativistic enthalpy density. This bound implies that, in ideal gases with gapped non-hydrodynamic sector, the shear viscosity is necessarily finite (if one neglects stochastic fluctuations \cite{KovtunStickiness2011}). For gapless systems, instead, $\eta$ might diverge, although this does not always happen (indeed, $\lambda \varphi^4$ kinetic theory is known to have finite $\eta$ \cite{Rocha:2023hts}). To better understand the relationship between gaplessness and divergence of $\eta$, consider the Kubo formula for the shear viscosity of a gas \cite{Denicol_Relaxation_2011} in the relaxation-time-approximation (with energy-dependent relaxation time $\tau_p$):
\begin{equation}
\eta= \dfrac{1}{T} \int \dfrac{d^3p}{(2\pi)^3} f^{\text{eq}}_p \bigg( \dfrac{p^1 p^2}{p^0}\bigg)^2 \tau_p \, .
\end{equation}
It can be seen that, for $\eta$ to diverge, $\tau_p$ should diverge fast enough to win over the exponential decay of $f^{\text{eq}}_p=e^{(\mu-p^0)/T}$. This, in turn, happens if the total scattering cross-section $\sigma(s) \propto \tau^{-1}_p$ decays fast enough. For example, $\eta$ is found to diverge whenever $\sigma(s) \sim \exp(-s^\alpha)$, with $\alpha>1$. Gases of this type do not admit a rigorous Navier-Stokes description. It remains an open question under which conditions a gapless system admits a convergent derivative expansion.

Through Theorem \ref{theo2}, we gave a sufficient condition for the existence of the gap (i.e., for $\tau_g<\infty$). Namely, the total collision cross-section should not approach zero for any $s$. This implies, for example, that a gas with constant cross-section is gapped. However, if one follows our discussion of section \ref{thegappone}, the lower bound \eqref{TheBound} was used only once, in the proof that the energy-dependent relaxation time \eqref{relxa} is bounded above by a positive constant. The same fact is known to hold for all ``hard interactions'' \cite[Theorem 4.1.A]{Dudynski1988}, namely for all differential cross sections such that
\begin{equation}
\upvarsigma(s,\theta) > \dfrac{A (s{-}4m^2)^{(a+1)/2}}{B+(s{-}4m^2)^{1/2}}\sin^{b}\theta \, ,
\end{equation}
where $A,B,a,b$ are all constants, with $A$ and $B$ positive, while $0\leq a < b{+}2$. Thus, for such interactions, the non-hydrodynamic sector is also gapped, and the results of Theorem \ref{theo1} hold, in agreement with the discussion in \cite{Dudynski1989}.

\section*{Acknowledgements}

This work was supported by a Vanderbilt's Seeding Success Grant. I thank J. Noronha for reading the manuscript and providing useful feedback.

\appendix

\section{Properties of the linearised Boltzmann operator}\label{proproiuzzione}

\subsection{Linearizing Boltzmann's equation}

We recall that, if $f(x^\alpha,p^j)$ is the kinetic distribution function of an ideal relativistic gas of classical particles, the relativistic Boltzmann equation takes the form \cite{Groot1980RelativisticKT}
\begin{equation}\label{boltzmann}
    p^\mu \partial_\mu f_p =\dfrac{1}{2} \int dQ dQ' dP' \, W_{pp'\leftrightarrow qq'} \, (f_q f_{q'}-f_p f_{p'}) \, ,
\end{equation}
where $dP=d^3p/[(2\pi)^3 p^0]$ is the Lorentz-invariant integration measure, and $W_{pp'\leftrightarrow qq'}$ is the transition rate, which has some natural symmetries: $    W_{pp'\leftrightarrow qq'}=W_{qq'\leftrightarrow pp'}=W_{p'p\leftrightarrow qq'}$.
Fixed the uniform state of global thermodynamic equilibrium $f_p^{\text{eq}}=e^{(\mu-p^0)/T}$, the conservation laws (in particular $p+p'=q+q'$) imply that
\begin{equation}\label{equilibrium}
    f^{\text{eq}}_p f^{\text{eq}}_{p'} = f^{\text{eq}}_q f^{\text{eq}}_{q'}
\end{equation}
whenever the transition $pp'{\leftrightarrow} qq'$ is allowed (namely when $W_{pp'\leftrightarrow qq'} \neq 0$). It follows that $f_p^{\text{eq}}$ is an exact solution of \eqref{boltzmann}. We are interested in studying small perturbations around this state. As we said in the main text, we write a generic distribution function as $f_p=f^\text{eq}_p(1+\phi_p)$, where $\phi_p$ is the relative deviation of the state $f_p$ from the global thermodynamic equilibrium state $f^{\text{eq}}_p$. Linearising \eqref{boltzmann} in $\phi_p$, and invoking \eqref{equilibrium}, we obtain:
\begin{equation}\label{linBoltzmann}
    p^\mu \partial_\mu \phi_p = \dfrac{1}{2} \int dQ dQ' dP' \, W_{pp'\leftrightarrow qq'} \,  f^{\text{eq}}_{p'}(\phi_q+\phi_{q'}-\phi_p-\phi_{p'}) \, .
\end{equation}
Recalling that $I=-(p^0)^{-1}L$, we have
\begin{equation}
    \begin{split}
& I[\phi_p] =  \dfrac{1}{2} \int \dfrac{d^3 q}{(2\pi)^3} \dfrac{d^3 q'}{(2\pi)^3} \dfrac{d^3 p'}{(2\pi)^3}  \Bar{W}_{pp'\leftrightarrow qq'} \,  f^{\text{eq}}_{p'}(\phi_p+\phi_{p'}-\phi_q-\phi_{q'}) \,  ,\\
& \text{with } \, \, \Bar{W}_{pp'\leftrightarrow qq'} \equiv \dfrac{ W_{pp'\leftrightarrow qq'}}{p^0 p'^0 q^0 q'^0} \, . \\
    \end{split}
\end{equation}
The new quantity $\Bar{W}_{pp'\leftrightarrow qq'}$ may be viewed as a ``non-covariant'' transition rate, and it obeys the same symmetries as $W_{pp'\leftrightarrow qq'}$, namely $    \Bar{W}_{pp'\leftrightarrow qq'}=\Bar{W}_{qq'\leftrightarrow pp'}=\Bar{W}_{p'p\leftrightarrow qq'}$. 

\subsection{Thermodynamic properties}\label{thermoprop}

The goal is to prove the relations in equation \eqref{propertiesfgofI}. Clearly, $I[1]=0$, because $\phi_p+\phi_{p'}-\phi_q-\phi_{q'}=1+1-1-1=0$. We find that $I[p^\nu]$ vanishes because $\Bar{W}_{pp'\leftrightarrow qq'}\propto \delta^4(p+p'-q-q')$, so that $\phi_p+\phi_{p'}-\phi_q-\phi_{q'}=p^\nu+p'^\nu-q^\nu-q'^\nu=0$. The proof that $I[\phi_p]$ vanishes \textit{only if} $\phi_p=a+b_\nu p^\nu$ is more elaborate. We only provide a reference \cite[Ch IX, \S 1]{Groot1980RelativisticKT}.
 
Let us now prove the Hermiticity of $I$. To this end, we first split $(\phi_p,I\psi_p)$ as the sum of four parts:
\begin{equation}\label{laquarta}
\begin{split}
(\phi_p,I\psi_p) = {}& \dfrac{1}{2} \int \dfrac{d^3 q}{(2\pi)^3} \dfrac{d^3 q'}{(2\pi)^3} \dfrac{d^3 p}{(2\pi)^3}\dfrac{d^3 p'}{(2\pi)^3}  \Bar{W}_{pp'\leftrightarrow qq'} \,  f^{\text{eq}}_{p} f^{\text{eq}}_{p'} \phi_p^* (\psi_p+\psi_{p'}-\psi_q-\psi_{q'}) \\
= {}& \dfrac{1}{2} \int \dfrac{d^3 q}{(2\pi)^3} \dfrac{d^3 q'}{(2\pi)^3} \dfrac{d^3 p}{(2\pi)^3}\dfrac{d^3 p'}{(2\pi)^3}  \Bar{W}_{pp'\leftrightarrow qq'} \,  f^{\text{eq}}_{p} f^{\text{eq}}_{p'} \phi_p^* \psi_p \\
+{}& \dfrac{1}{2} \int \dfrac{d^3 q}{(2\pi)^3} \dfrac{d^3 q'}{(2\pi)^3} \dfrac{d^3 p}{(2\pi)^3}\dfrac{d^3 p'}{(2\pi)^3}  \Bar{W}_{pp'\leftrightarrow qq'} \,  f^{\text{eq}}_{p} f^{\text{eq}}_{p'} \phi_p^* \psi_{p'} \\
-{}&\dfrac{1}{2} \int \dfrac{d^3 q}{(2\pi)^3} \dfrac{d^3 q'}{(2\pi)^3} \dfrac{d^3 p}{(2\pi)^3}\dfrac{d^3 p'}{(2\pi)^3}  \Bar{W}_{pp'\leftrightarrow qq'} \,  f^{\text{eq}}_{p} f^{\text{eq}}_{p'} \phi_p^* \psi_q \\
-{}& \dfrac{1}{2} \int \dfrac{d^3 q}{(2\pi)^3} \dfrac{d^3 q'}{(2\pi)^3} \dfrac{d^3 p}{(2\pi)^3}\dfrac{d^3 p'}{(2\pi)^3}  \Bar{W}_{pp'\leftrightarrow qq'} \,  f^{\text{eq}}_{p} f^{\text{eq}}_{p'} \phi_p^* \psi_{q'} \, . \\
\end{split}
\end{equation}
Now, we leave the first term as it is. In the second term, we use the symmetry $\Bar{W}_{pp'\leftrightarrow qq'} \,  f^{\text{eq}}_{p} f^{\text{eq}}_{p'}=\Bar{W}_{p'p\leftrightarrow qq'} \,  f^{\text{eq}}_{p'} f^{\text{eq}}_{p}$, and we rename the (dummy) integration variables as follows: $(qq'pp')\rightarrow (qq'p'p)$. In the third term, we use the symmetry $\Bar{W}_{pp'\leftrightarrow qq'} \,  f^{\text{eq}}_{p} f^{\text{eq}}_{p'}=\Bar{W}_{qq'\leftrightarrow pp'} \,  f^{\text{eq}}_{q} f^{\text{eq}}_{q'}$, and we rename the integration variables as follows: $(qq'pp')\rightarrow (pp'qq')$. In the last term, we use the symmetry $\Bar{W}_{pp'\leftrightarrow qq'} \,  f^{\text{eq}}_{p} f^{\text{eq}}_{p'}= \Bar{W}_{q'q\leftrightarrow p'p} \,  f^{\text{eq}}_{q'} f^{\text{eq}}_{q}$, and we rename the variables as follows: $(qq'pp')\rightarrow (p'pq'q)$. Taking the complex conjugate of \eqref{laquarta}, we obtain
\begin{equation}\label{laquarta2}
\begin{split}
(\phi_p,I\psi_p)^*
= {}& \dfrac{1}{2} \int \dfrac{d^3 q}{(2\pi)^3} \dfrac{d^3 q'}{(2\pi)^3} \dfrac{d^3 p}{(2\pi)^3}\dfrac{d^3 p'}{(2\pi)^3}  \Bar{W}_{pp'\leftrightarrow qq'} \,  f^{\text{eq}}_{p} f^{\text{eq}}_{p'} \psi_p^* \phi_p \\
+{}& \dfrac{1}{2} \int \dfrac{d^3 q}{(2\pi)^3} \dfrac{d^3 q'}{(2\pi)^3} \dfrac{d^3 p}{(2\pi)^3}\dfrac{d^3 p'}{(2\pi)^3}  \Bar{W}_{pp'\leftrightarrow qq'} \,  f^{\text{eq}}_{p} f^{\text{eq}}_{p'} \psi_p^* \phi_{p'} \\
-{}&\dfrac{1}{2} \int \dfrac{d^3 q}{(2\pi)^3} \dfrac{d^3 q'}{(2\pi)^3} \dfrac{d^3 p}{(2\pi)^3}\dfrac{d^3 p'}{(2\pi)^3}  \Bar{W}_{pp'\leftrightarrow qq'} \,  f^{\text{eq}}_{p} f^{\text{eq}}_{p'} \psi_p^* \phi_q \\
-{}& \dfrac{1}{2} \int \dfrac{d^3 q}{(2\pi)^3} \dfrac{d^3 q'}{(2\pi)^3} \dfrac{d^3 p}{(2\pi)^3}\dfrac{d^3 p'}{(2\pi)^3}  \Bar{W}_{pp'\leftrightarrow qq'} \,  f^{\text{eq}}_{p} f^{\text{eq}}_{p'} \psi_p^* \phi_{q'}  \\
= {}& \dfrac{1}{2} \int \dfrac{d^3 q}{(2\pi)^3} \dfrac{d^3 q'}{(2\pi)^3} \dfrac{d^3 p}{(2\pi)^3}\dfrac{d^3 p'}{(2\pi)^3}  \Bar{W}_{pp'\leftrightarrow qq'} \,  f^{\text{eq}}_{p} f^{\text{eq}}_{p'} \psi_p^* (\phi_p+\phi_{p'}-\phi_q-\phi_{q'})=(\psi_p,I\phi_p) \, , \\
\end{split}
\end{equation}
thereby proving that $I$ is symmetric.
The last fact that we need to prove is that $(\phi_p,I\phi_p)\geq 0$. This can be done directly, with a strategy similar to the above. In fact, we have the following equalities:
\begin{equation}\label{finione}
\begin{split}
(\phi_p,I\phi_p) = {}& \dfrac{1}{2} \int \dfrac{d^3 q}{(2\pi)^3} \dfrac{d^3 q'}{(2\pi)^3} \dfrac{d^3 p}{(2\pi)^3}\dfrac{d^3 p'}{(2\pi)^3}  \Bar{W}_{pp'\leftrightarrow qq'} \,  f^{\text{eq}}_{p} f^{\text{eq}}_{p'} \phi_p^* (\phi_p+\phi_{p'}-\phi_q-\phi_{q'}) \\
= {}& \dfrac{1}{2} \int \dfrac{d^3 q}{(2\pi)^3} \dfrac{d^3 q'}{(2\pi)^3} \dfrac{d^3 p}{(2\pi)^3}\dfrac{d^3 p'}{(2\pi)^3}  \Bar{W}_{pp'\leftrightarrow qq'} \,  f^{\text{eq}}_{p} f^{\text{eq}}_{p'} \phi_{p'}^* (\phi_p+\phi_{p'}-\phi_q-\phi_{q'}) \\
= {}& \dfrac{1}{2} \int \dfrac{d^3 q}{(2\pi)^3} \dfrac{d^3 q'}{(2\pi)^3} \dfrac{d^3 p}{(2\pi)^3}\dfrac{d^3 p'}{(2\pi)^3}  \Bar{W}_{pp'\leftrightarrow qq'} \,  f^{\text{eq}}_{p} f^{\text{eq}}_{p'} (-\phi_q^*) (\phi_p+\phi_{p'}-\phi_q-\phi_{q'}) \\
= {}& \dfrac{1}{2} \int \dfrac{d^3 q}{(2\pi)^3} \dfrac{d^3 q'}{(2\pi)^3} \dfrac{d^3 p}{(2\pi)^3}\dfrac{d^3 p'}{(2\pi)^3}  \Bar{W}_{pp'\leftrightarrow qq'} \,  f^{\text{eq}}_{p} f^{\text{eq}}_{p'} (-\phi_{q'}^*) (\phi_p+\phi_{p'}-\phi_q-\phi_{q'}) \, . \\
\end{split}
\end{equation}
In each line, we performed a change of variables similar to those used for converting \eqref{laquarta} into \eqref{laquarta2}. Adding up all four lines in \eqref{finione}, we finally obtain the desired result:
\begin{equation}
(\phi_p,I\phi_p) =  \dfrac{1}{8} \int \dfrac{d^3 q}{(2\pi)^3} \dfrac{d^3 q'}{(2\pi)^3} \dfrac{d^3 p}{(2\pi)^3}\dfrac{d^3 p'}{(2\pi)^3}  \Bar{W}_{pp'\leftrightarrow qq'} \,  f^{\text{eq}}_{p} f^{\text{eq}}_{p'} |\phi_p+\phi_{p'}-\phi_q-\phi_{q'}|^2 \geq 0 \, . 
\end{equation}

\subsection{Behaviour under rotations}\label{ououritazia}

Let us prove that $\mathcal{R}$ commutes with $I$.
To this end, we need to invoke the invariance of the transition rate under rotations (namely the identity $\Bar{W}_{pp'\leftrightarrow qq'}=\Bar{W}_{\mathcal{R}p\mathcal{R}p'\leftrightarrow \mathcal{R}q \mathcal{R}q'}$) and the isotropy of the equilibrium distribution:
\begin{equation}\label{theorsiu}
\begin{split}
I  \mathcal{R} \phi_p={}& \dfrac{1}{2} \int \dfrac{d^3 q}{(2\pi)^3} \dfrac{d^3 q'}{(2\pi)^3} \dfrac{d^3 p'}{(2\pi)^3}  \Bar{W}_{pp'\leftrightarrow qq'} \,  f^{\text{eq}}_{p'}\big[(\mathcal{R}\phi)_p+(\mathcal{R}\phi)_{p'}-(\mathcal{R}\phi)_q-(\mathcal{R}\phi)_{q'}\big] \\
   ={}& \dfrac{1}{2} \int \dfrac{d^3 q}{(2\pi)^3} \dfrac{d^3 q'}{(2\pi)^3} \dfrac{d^3 p'}{(2\pi)^3}  \Bar{W}_{pp'\leftrightarrow qq'} \,  f^{\text{eq}}_{p'}\big[\phi_{\mathcal{R}p}+\phi_{\mathcal{R}p'}-\phi_{\mathcal{R}q}-\phi_{\mathcal{R}q'}\big] \\
   ={}& \dfrac{1}{2} \int \dfrac{d^3 q}{(2\pi)^3} \dfrac{d^3 q'}{(2\pi)^3} \dfrac{d^3 p'}{(2\pi)^3}  \Bar{W}_{\mathcal{R}p\mathcal{R}p'\leftrightarrow \mathcal{R}q \mathcal{R}q'} \,  f^{\text{eq}}_{\mathcal{R}p'}\big[\phi_{\mathcal{R}p}+\phi_{\mathcal{R}p'}-\phi_{\mathcal{R}q}-\phi_{\mathcal{R}q'}\big] \\
   ={}& \dfrac{1}{2} \int \dfrac{d^3 q}{(2\pi)^3} \dfrac{d^3 q'}{(2\pi)^3} \dfrac{d^3 p'}{(2\pi)^3}  \Bar{W}_{\mathcal{R}pp'\leftrightarrow qq'} \,  f^{\text{eq}}_{p'}\big[ \phi_{\mathcal{R}p}+\phi_{p'}-\phi_q-\phi_{q'}\big]= (I\phi)_{\mathcal{R}p}=\mathcal{R}I\phi_p \, . \\
   \end{split}
\end{equation}
In the last line, we redefined the integration variables as follows: $(q,q',p') \rightarrow (\mathcal{R}^{-1}q,\mathcal{R}^{-1}q',\mathcal{R}^{-1}p')$. The expression ``$(I\phi)_{\mathcal{R}p}$'' means that we evaluate the function $I\phi$ (i.e. the output of $I$ when acting on $\phi$) at momentum $\mathcal{R}p$. Since the reasoning in equation \eqref{theorsiu} is valid for all $\phi_p$ in $\mathcal{H}$, we conclude that $I \mathcal{R}=\mathcal{R}I$, completing our proof.

\bibliography{Biblio}

\label{lastpage}

\end{document}